\documentclass[twocolumn,aps,floatfix,superscriptaddress,prl,showpacs]{revtex4}

\usepackage{graphicx}
\usepackage{amsmath,amssymb,amsfonts}
\usepackage{bm}
\usepackage{bibentry, natbib}
\usepackage[T1]{fontenc}
\usepackage[english]{babel}
\usepackage[latin1]{inputenc}
\usepackage{ifthen}
\usepackage{footnote}
\usepackage{subfigure}
\usepackage{setspace}

\newcommand{\be}{\begin{equation}}
\newcommand{\ee}{\end{equation}}
\newcommand{\rmd}{{\rm{d}}}
\newcommand{\rme}{{\rm e}}
\newcommand{\re}{{\rm Re\,}}
\newcommand{\reff}[1]{(\ref{#1})}
\newcommand{\cf}{{\mathcal{F}}}

\newcommand{\lext}{\ell_{\rm ext}}

\newcommand{\ba}{\begin{align}}
\newcommand{\ea}{\end{align}}

\addto\captionsenglish{}

\begin{document}

%\setstretch{2.5}

\title{Large deviations and universality %of the work done
in quantum quenches}
\author{Andrea Gambassi}
\affiliation{SISSA -- International School for Advanced Studies, via Bonomea 265, 34136 Trieste, Italy}
\affiliation{INFN -- Istituto Nazionale di Fisica Nucleare, sezione di Trieste}
\author{Alessandro Silva} 
\affiliation{SISSA -- International School for Advanced Studies, via Bonomea 265, 34136 Trieste, Italy}
\affiliation{Abdus Salam ICTP, Strada Costiera 11, 34151 Trieste, Italy}

\date{\today}

\begin{abstract}
We study the large deviations statistics of the intensive work done by changing globally a control parameter in a thermally isolated quantum many-body system. We show that, upon approaching a critical point, large deviations well below the mean work display universal features related to the critical Casimir effect in the corresponding classical system.
Large deviations well above the mean are, instead, of quantum nature and not captured by the quantum-to-classical correspondence. For a bosonic system we show that in this latter 
regime a transition from exponential to power-law statistics, analogous to the equilibrium Bose-Einstein condensation, may occur depending on the parameters of the quench and on the spatial dimensionality.  
\end{abstract}

%\pacs{}

\maketitle

\emph{Introduction --} %
Recent experimental progresses in the physics of trapped ultracold atomic gases have stimulated a growing interest in the non-equilibrium behavior of thermally isolated quantum many-body systems \cite{PSSV-11}. 
A number of aspects are presently being investigated experimentally, ranging from the propagation of correlations
after quenches~\cite{Cheneau2012} to relaxation and pre-thermalization inferred from the statistical fluctuations of the interference contrast of split condensates~\cite{Gring2012}. 
On the theoretical side, a compelling  
issue under investigation is that of the role played by universality in the non-equilibrium dynamics~\cite{PSSV-11}, since predictions independent of microscopic details make the comparison with 
experiments a particularly stringent test. 
Universal behavior can be investigated  by studying either the time dependence of correlation functions~\cite{CC-06,CG-11}, in particular close to criticality, or their statistical fluctuations~\cite{Kitagawa2011}.  In this context, a number of studies have focused on macroscopic, thermodynamic variables such as work~\cite{S-08, GS-11, Heyl2012, Bunin2012,Smacchia2012, HPK-12} and entropy~\cite{Polkovnikov2010},
exploring the emergence of universality in their statistical fluctuations.
 
Statistical fluctuations are known to provide insight into the physics of classical equilibrium and non-equilibrium systems~\cite{T-09}. 
The statistics of macroscopic \emph{extensive} variables exhibits a first, obvious form of universality associated to typical, "small" fluctuations, which is however rather insensitive to the underlying properties of the system~\cite{T-09}.
Indeed, as the mean of a generic extensive quantity $W_N$ (e.g., the magnetization in a spin system) grows proportionally to the number $N$ of degrees of freedom,  
the one of the associated \emph{intensive} variable $w_N \equiv W_N/N$ (i.e., the magnetization per unit volume)
approaches a finite value $\bar w$. 
The central limit theorem (when applicable) suggests that the typical fluctuations of $w_N$ are suppressed $\sim 1/N^{1/2}$ and have a Gaussian distribution around $\bar w$. 
On the other hand, large fluctuations, though rare, are capable of probing the specific details of the
physical system~\cite{T-09} and they might provide valuable information on its universal behavior. 
In order for a large fluctuation to occur, an extensively large number of microscopic fluctuating variables (i.e., the spin, in our example) has to deviate significantly from their corresponding means \cite{footnote1} and this
happens with a probability which is exponentially small in the size $N$. 
Accordingly,  for large $N$, one expects $w_N$ to be distributed according to a probability density $p(w) \sim \exp[-N I(w)]$, where the so-called \emph{rate function} $I(w)$ is non-negative,
%(as required by the normalization of $p$), 
vanishes for $w = \bar w$ \cite{footnote2}, and characterizes the statistics of both large deviations and Gaussian fluctuations. 

Here we show that the statistics of large deviations of the intensive work $w$ done during a global quench of a thermally isolated quantum many-body system provides 
insight into its universal properties.
For a global quench one heuristically expects $p(w)$ to feature
a prominent Gaussian peak centered at a finite mean $\bar{w}$. By focusing
on the tails of this distribution, we demonstrate that there is a clear distinction between large deviations well below ($w \ll \bar{w}$) and well above  ($w \gg \bar{w}$) the mean. The former are determined by the excess free energy 
$f_{ex}$ of the $d+1$ dimensional classical correspondent in a film~\cite{GS-11,CG-11}
and acquire universal features close to a possible critical point. 
The latter, instead, are genuinely quantum features, beyond the quantum-to-classical correspondence, which may, however, maintain some tracts of universality. 
Our analysis encompasses as examples
the cases of quenches in the quantum Ising chain~\cite{S-08} and in a free bosonic system~\cite{GSS-12}.
In addition, we show that, depending on the space dimensionality,   the large deviation statistics of bosonic systems displays a so-called \emph{condensation transition} (see, e.g., Ref.~\cite{MK-10}), analogous to the Bose-Einstein condensation.

\emph{Statistics of the work --} Consider a quantum system with $N$ interacting degrees of freedom and Hamiltonian $H(g)$. The extensive work $W_N$  performed on the system during the %(instantaneous) 
quench $g_0\to g$ is determined by the initial state, typically the ground state $|\Psi^{g_0}_0\rangle$ of $H(g_0)$, and by the eigenvalues $E_{n\ge 0}^{g}$ and 
eigenvectors $|\Psi_n^g\rangle$ of the post-quench Hamiltonian $H(g)$. In particular, 
$W_N$ is a stochastic variable with probability density~\cite{Kurchan2000,Talkner2007}
\be
p(W_N) = \sum_{n\ge 0} |\langle \Psi_n^g| \Psi_0^{g_0} \rangle |^2\delta(W_N - [E_n^g-E_0^{g_0}]),
\label{eq:pdfWN}
\ee
where $E_0^g$ indicates generically the extensive ground-state energy of $H(g)$.  As $p$ vanishes identically for $W_N$ below   
$E_0^g-E_0^{g_0}$, we refer $W_N$ to this threshold
so that $W_N\ge 0$. 

The   
probability $p(W_N)$ can be conveniently studied 
via its moment generating function
\be
G(s) \equiv \langle \rme^{- s W_N} \rangle,
\label{eq:mgf}
\ee
which, for $N\to\infty$, 
exists in the complex half-plane containing $\re s \ge 0$ (with possible zeros, see, e.g., Ref.~\cite{HPK-12}). 
For later purposes, we distinguish here a class A of systems
in which $W_N$ for large but finite $N$ cannot exceed a certain extensive threshold $Nw_M$ from the class B whithin which $W_N$ can assume arbitrarily large values.
%
%%%% cases A and B:
Generically, in the former class, $G(s)$ is defined for all $s\in \mathbb{R}$ with $G(s)\simeq \rme^{-s N w_M}$ for $s\to -\infty$, whereas in the latter $G(s)$ is defined only for $s>-\bar s < 0$ with a generic singularity in its derivative at $-\bar s$. The quantum Ising chain in a transverse field  and the free bosonic field belong to classes A and B, respectively. 
%
%

%
%
%%%%%%%%%%%%%%%%%%%%%%%%%%%%%
\begin{figure}[t!]
\begin{center}
\includegraphics[scale=0.32]{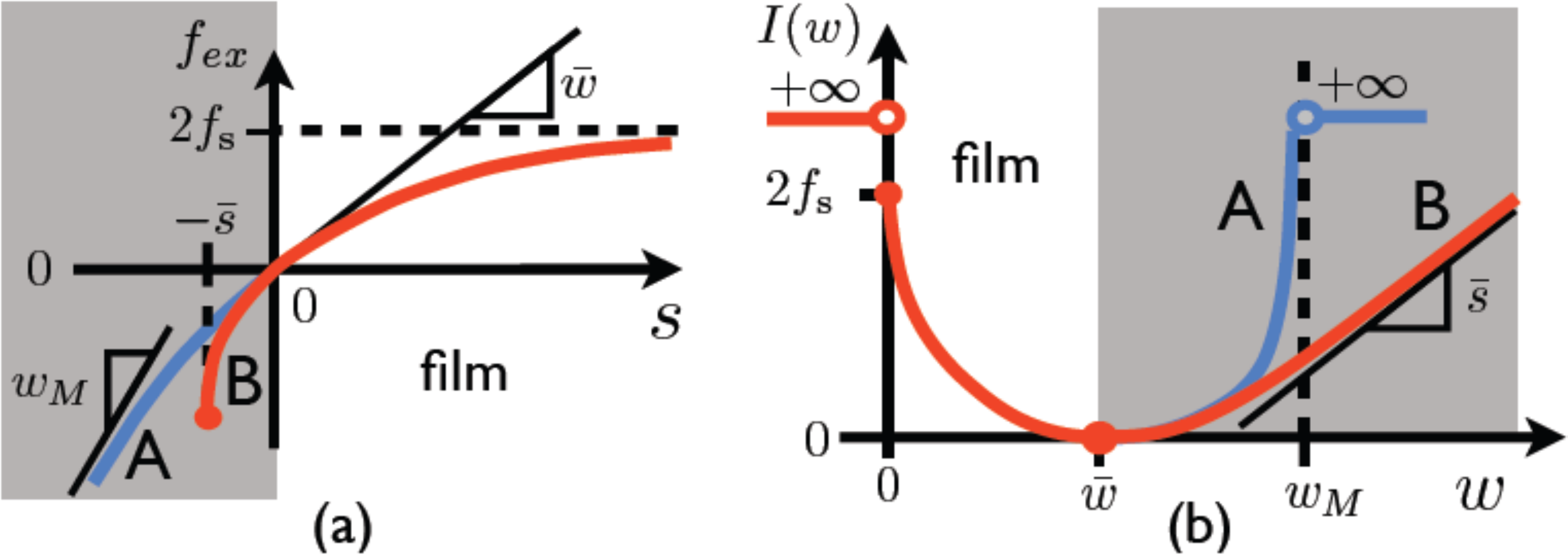}
\end{center}
\caption{(a) Sketch of the excess free energy density $f_{ex}(s)$ and (b) of the corresponding rate function $I(w)$ for classes A (blue) and B (red) discussed in the main text. The gray area highlights the range of variables for which $f_{ex}$ does not have a thermodynamic interpretation.}
\label{fig:LF}
\end{figure}
%%%%%%%%%%%%%%%%%%%%%%%%%%%%%
%

The \emph{quantum} to \emph{classical} correspondence 
allows us to interpret the moment generating function
$G(s)$ for $s>0$ as the \emph{partition function} of a \emph{classical} system %in $d+1$ dimensions 
in a film geometry~\cite{GS-11}. Indeed, Eq.~\reff{eq:pdfWN} implies
\be
G(s) = \langle \Psi_0^{g_0}| \rme^{-s [H(g) - E_0^g]}|\Psi_0^{g_0}\rangle,
\label{eq:def-G}
\ee
where $ \langle \Psi_0^{g_0}| \rme^{-s H(g)}|\Psi_0^{g_0}\rangle \equiv Z_{N\times s}$ is in fact such a
partition function of the classical $d+1$-dimensional system with transfer matrix $\rme^{-H(g)}$ corresponding to the \emph{quantum} Hamiltonian $H(g)$, in a film geometry with transverse "surface" area $N$, "thickness" $s$ and equal boundary conditions set by $|\Psi_0^{g_0}\rangle$. On the basis of $Z_{N\times s}$ one naturally defines the free energy $\cf_{N\times s} \equiv - \ln Z_{N\times s}$ per $k_BT$, where $T$ is the temperature of the corresponding classical system, which depends on the parameters of $H(g)$. 
In terms of the classical system, the variable $s$ in Eq.~\reff{eq:mgf} is the distance between the two confining surfaces which we assume to have a large transverse area $N$.
Upon increasing $s$, the free energy density per unit area $f$  
decomposes in decreasing powers of $s$ as \cite{FSS}
\be
f \equiv N^{-1}\cf_{N\times s} \simeq  s  f_b + 2 f_s + {\rm corr.},
\label{eq:dec-F}
\ee
where $f_b= \lim_{N,s\to\infty }\cf_{N\times s}/(N s)$ is the bulk free energy density and $f_s$ is the surface free energy density, 
i.e., the energy cost for introducing separately each single boundary into the otherwise bulk system. The corrections "corr." in Eq.~\reff{eq:dec-F}  vanish for $s\to\infty$.
In order to separate the effects of confinement from the bulk behavior, one usually introduces the so-called \emph{excess}  free energy density per unit area $f_{ex} \equiv f - s f_b$, which
plays a fundamental role in what follows and becomes independent of $N$ 
in the limit of large $N$ considered hereafter.  
In terms of the quantum system, one finds from Eqs.~\reff{eq:def-G} and \reff{eq:dec-F}, that $f_b = E_0^g/N$, $f_s = -(\ln |\langle \Psi_0^{g_0}|\Psi_0^g\rangle |)/N$ \cite{GS-11} 
and therefore 
\be
G(s) = \rme^{-N f_{ex}(s)}.
\label{eq:scaled-cgf}
\ee
For $s<0$, $f_{ex}$ is defined in terms of $G(s)$ by this equations and it lacks its thermodynamic interpretation.

\emph{Large deviations and universality --} %
Equation~\reff{eq:scaled-cgf} is crucial for understanding 
the emergence of universality in the large deviations statistics of the intensive work $w_N=W_N/N$. 
In fact, 
its distribution $p(w)$ for $N\to\infty$ can be determined by a saddle-point approximation of the inverse Laplace transform of $G(s)$, which actually provides a heuristic derivation of the G\"artner-Ellis theorem \cite{T-09}. In particular, Eq.~\reff{eq:scaled-cgf} implies that $p(w)$ has the form $\sim \exp[-N I(w)]$, where the \emph{rate function} $I(w)$ is %given by 
the Legendre-Fenchel transform of $f_{ex}(s)$ (and viceversa, under certain assumptions~\cite{T-09})
\be
I(w) = - \inf_{s\in\mathbb{R}} \, \left\{s w - f_{ex}(s)\right\},
\label{eq:LF}
\ee
in which the infimum is taken within the domain of definition of $f_{ex}(s)$ and $G(s)$. 

The generic features of $p(w)$ 
can now be inferred from Eqs.~\reff{eq:scaled-cgf} and \reff{eq:LF}.
First of all note that the excess free energy is such that $f_{ex}(0)=0$ and $f_{ex}'(0)=\bar w$.
Most importantly $f_{ex}(s)$ is a concave function of $s$ \cite{T-09} which approaches 
$2 f_s$ for $s\to+\infty$.
Figure \ref{fig:LF} provides a sketch of $f_{ex}(s)$ and the corresponding $I(s)$ for the two classes A and B introduced above.
The last two properties imply the existence of a threshold in $p(w)$: the infimum in Eq.~\reff{eq:LF} for $w<0$ is $-\infty$  and consequently $p(w<0)=0$. The behavior of $I(w)$ close to the threshold $w\gtrsim 0$, instead, is determined by the one of $f_{ex}(s)$ for $s\to+\infty$ and in particular $I(0) =  2 f_s > 0$, while the approach to it 
is determined by the corrections $f_{ex}-2 f_s$ in Eq.~\reff{eq:dec-F}. 

The universality of these finite-size corrections %to $f_{ex}(s)$ 
close to critical points \cite{G-09} carries over into  
the large deviation statistics of $p(w)$ for $w \ll \bar{w}$. Indeed,  if the post-quench Hamiltonian $H(g)$ is close to a quantum critical point the finite-size corrections $f_{ex}-2 f_s$ to the free energy density of the (near-critical) classical $d+1$-dimensional system, which are responsible for the so-called critical Casimir effect \cite{G-09}, take the universal scaling form $s^{-d}\Theta(s/\xi)$ for $s\gg a$, where $\xi \gg a$ is the correlation length and $a$ some microscopic length scale. The scaling function $\Theta$ is \emph{universal} in the sense of critical phenomena \cite{G-09}, as it depends only on the universality class of the classical critical point. In addition, due to the presence of the boundaries, $\Theta$ depends on their \emph{surface universality class} \cite{D-86} or, equivalently, on which among the few effective boundary (i.e., initial) states $\{|\Psi^*_i\rangle\}_i$, $|\Psi_0^{g_0}\rangle$ flows to %under scaling 
as the critical point is approached. 
Once the scaling function $\Theta$ is known, the rate function is calculated via Eq.~\reff{eq:LF}. 
In particular, if the post-quench Hamiltonian is critical, then $\xi=\infty$ and 
\be
I(w\lesssim \Delta ) =  2 f_s - \frac{d+1}{d}\Delta \left(\frac{w}{\Delta}\right)^{d/(d+1)} + \dots
\label{eq:Iw0}
\ee
with $\Delta = d |\Theta(0)|$.
While $f_s$ and the possible corrections depend on the specific parameters of the initial state,
the leading dependence of $I(w)$ on $w$ is universal and non-analytic.  
In the case of finite but large $\xi$, the approach to the value $2 f_s$ is eventually controlled by $\Theta(x\gg 1) = {\mathcal C} x^a\rme^{-b x}$ where the universal constants $a$, $b$, and ${\mathcal C}$ depend, along with $\Theta$, on the bulk and surface universality class of the transition, and they are known for a variety of universality classes \cite{G-09} (e.g.,  $a=-1/2$, $b=2$ for a quench of the quantum Ising chain within the same phase~\cite{GS-11}). 
In this case, one finds %at the leading order 
$I(w\ll \xi^{-(d+1)}) \simeq 2 f_s - (\xi/b)w\ln w^{-1}$ but with significant logarithmic corrections. Notice that though this leading order is independent of $a$, a study of the 
rate function $I(w)$ for larger values of $w$ provides information on the complete scaling function $\Theta$, and hence on the corresponding boundary universality class.
Accordingly, not only the edge singularities of %statistics of the 
the extensive work $W_N$ %close to the threshold 
studied in Ref.~\cite{GS-11} are determined by universal features of the system and of the quench, % which it is subject to, 
but also the large deviations of the intensive variable $w_N$ display universal properties in their rate function $I(w)$ close to the threshold~\cite{NotaBene}. 

In order to illustrate the discussion above we focus on a free bosonic theory
described by a Hamiltonian diagonalizable in independent momentum modes
\be
\label{ham1}
H(m) = \int\!\!\frac{\rmd^d k}{(2\pi)^d} \left( \frac{1}{2} \pi_{k}\pi_{-k} + \frac{1}{2} \omega^{2}_{k} \phi_{k}\phi_{-k} \right),
\ee
where $[\phi_k,\pi_{k'}]= i\delta_{k,k'}$
and the integral runs over the first Brillouin zone $|k_i|<\pi$.  
We assume a relativistic dispersion relation $\omega_k(m)=\sqrt{k^2+m^2}$ and consider quenches of the mass from $m_0$ to $m$ \cite{CC-06,GSS-12}. This Hamiltonian captures the low-energy properties of a number of physical systems, including the ideal harmonic chain, interacting fermions and bosons in one dimension~\cite{Cazalilla2011}, and it models the relative phase fluctuations of split one-dimensional condensates~\cite{Gring2012}.
%
%The Hamiltonian 
$H(m)$ has a critical point at $m=0$ and the corresponding classical theory is that of a Gaussian field 
$\varphi$ in $d+1$ spatial dimensions and mass $m$. 
The quench is characterized by $\lambda_k \equiv [\omega_k(m_0)-\omega_k(m)]/[\omega_k(m_0)+\omega_k(m)]$ and from Eqs.~\reff{eq:pdfWN}, \reff{eq:mgf}, and \reff{eq:scaled-cgf} one finds \cite{GSS-12}
\be
f_{ex}(s) = \frac{1}{2}\int\!\frac{\rmd^d k}{(2\pi)^d}\!\ln\left[ \frac{1-\lambda_k^2\rme^{-2\omega_k(m)s}}{1-\lambda_k^2}\right],
\label{eq:fex-HO}
\ee
which is defined for $s > - \bar s = \sup_k (\ln|\lambda_k|)/\omega_k(m)$ and, as anticipated, 
belongs to class B.  
This $f_{ex}$  
can be decomposed as in Eq.~\reff{eq:dec-F} and upon approaching the critical point $m=0$, i.e., for sufficiently large $\xi=m^{-1}$ and $s$, the correction $f_{ex}(s)-2 f_s$  takes the ($m_0$-independent) scaling form $s^{-d}\Theta_O(s/\xi)$ where $\Theta_O(x)$ is the scaling function of the critical Casimir effect for the classical field $\varphi$ with boundaries belonging to the so-called \emph{ordinary} surface universality class \cite{D-86}, corresponding to Dirichlet boundary conditions for $\varphi$. $\Theta_O$ can be read, e.g., in Eq.~(6.6) of Ref.~\cite{KD-92}. Accordingly, upon approaching the critical point, the ground state $|\Psi_0^{m_0}\rangle$ of $H(m_0)$ flows towards the fixed-point state $|\Psi_O^*\rangle$ corresponding to this surface universality class. 
However, as $s$ decreases, $f_{ex}(s)-2 f_s$ calculated from Eq.~\reff{eq:fex-HO} is no longer independent of $m_0$ and corrections to the critical Casimir term arise. These corrections are partly but effectively accounted for by changing $s \mapsto s + 2 \lext$ in the previous scaling form, where the so-called \emph{extrapolation length} $\lext$ \cite{D-86}
takes here the value $m_0^{-1}$.  This can be interpreted as the fact that the fixed-point Dirichlet boundary condition on $\varphi$ is effectively realized on confining surfaces which are at a distance $\lext$ from the boundaries of the film of thickness $s$,  resulting in an effective film of thickness $s + 2 \lext$ \cite{D-86,CC-06,CG-11}. For films sufficiently thick and critical, with $s,\xi\gg \lext$, this correction is unnecessary, while it becomes increasingly important as $\xi$, $s$, or $m_0$ decrease.
%
%
%
%
%%%%%%%%%%%%%%%%%%%%%%%%%%%%%
\begin{figure}[t!]
\begin{center}
\includegraphics[scale=0.41]{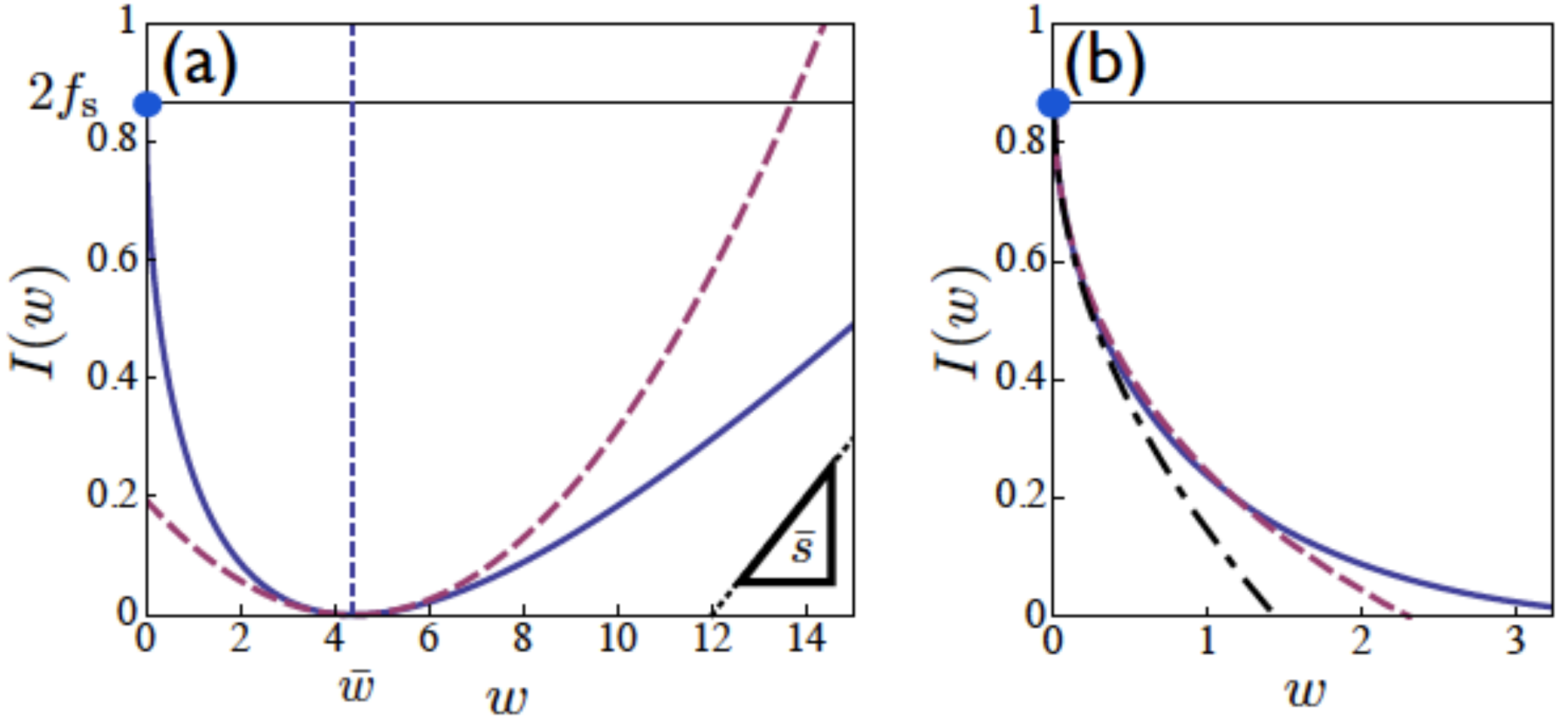}
\end{center}
\caption{Rate function $I(w)$ (solid line) of the work done on the lattice free bosonic theory in $d=1$ and unit lattice spacing, for a quench from $m_0 = 20$ to the critical point. % $m=0$. 
In panel (a) the dashed curve corresponds to the Gaussian distribution of small fluctuations around $\langle w \rangle = \bar w$. In panel (b), the dash-dotted curve is the prediction of Eq.~\reff{eq:Iw0} while the dashed curve accounts also for  a non-vanishing $\lext$.}
\label{fig:HO}
\end{figure}
%%%%%%%%%%%%%%%%%%%%%%%%%%%%%
%
%
%
Figure~\ref{fig:HO} presents the rate function $I(w)$ (solid line) in $d=1$ for a quench from a non-critical to the critical point. 
In panel (a) the thin vertical dashed line indicates the mean work $\bar w$. The
dashed curve, instead, provides %indicates 
the quadratic approximation of $I(w)$ around $w=\bar w$, which describes 
the Gaussian distribution of the small fluctuations. 
While additional features of this $I(w)$ are rationalized further below, we focus in panel (b) on 
the region of small $w$, where we expect universality to emerge.
The dash-dotted line correspond to Eq.~\reff{eq:Iw0}, with $\Theta_O(0) = -\Gamma(d) \zeta(d+1)/[(16\pi)^{d/2}\Gamma(d/2)]$ \cite{KD-92}.
This universal behavior sets in rather close to the threshold. However, the agreement between $I(w)$ 
and Eq.~\reff{eq:Iw0} extends to a wider range by accounting for the correction due to  
$\lext$ (dashed curve). 
The features displayed 
in Fig.~\ref{fig:HO} for $m=0$ carries over to the case %with 
$m \ll m_0$, which requires the knowledge of the full scaling function $\Theta_O(x)$. 
For a fixed value of $m$, instead, the corrections to the scaling behavior due to $\lext$ increase upon decreasing $m_0$ and eventually, after crossing the line $m=m_0$ of no quench, they lead to a change in the effective boundary state \cite{CG-11,GS-11} for $m_0 \to 0$.

\emph{Quantum regime and condensation -- } Let us now consider the case of large work $w \gg \bar{w}$.
Upon increasing $w$ %from $w<\bar{w}$, 
further away from the threshold, the value $s^*(w)$ of $s$ for which the infimum in Eq.~\reff{eq:LF} is attained --- and which satisfies $f_{ex}'(s^*(w))=w$ --- decreases and so does the thickness of the corresponding film. The behavior of such a film is expected to become increasingly dominated by its microscopic details, with a generic lack of universality even close to the critical point. Correspondingly $I$ decreases because $I'(w) = - s^*(w)$.  For $w = \bar w$,
$s^*=0$ and $I(w=\bar w)$ vanishes, while it grows again for $w>\bar{w}$, with $s^*(w)<0$ (see Fig.~\ref{fig:LF}).

The %behavior of the 
rate function for $w > \bar w$ is thus determined by
$f_{ex}(s)$ for $s<0$ ("negative" film thickness), which lacks a thermodynamic interpretation because the quantum-to-classical correspondence does not hold in this case. 
The qualitative behavior of $I(w>\bar w)$ depends crucially on the class the system belongs to.
In Fig.~\ref{fig:LF}  we report a sketch of (a) $f_{ex}(s)$ and (b) the associated rate function $I$ corresponding to classes A and B discussed above and characterized by (A) a bound (e.g., the quantum Ising chain) or (B) an unbound spectrum (e.g., free bosonic theory). 
In particular, in case A, $I(w)$ diverges upon approaching $w_M$, with $I(w>w_M) = +\infty$ as required by 
the fact that $p(w)$ vanishes above the intensive threshold $w_M$. In case B, instead, $I(w\to\infty)\simeq \bar s w$ and therefore $p(w\gg\bar w) \sim \rme^{-N\bar s w}$. This is seen in Fig.~\ref{fig:HO}(a),
though the asymptotic linear behavior for $w\gg 1$, with slope $\bar s$ (indicated by the dashed line) is actually approached only for rather large values of $w$. 
In general, 
$\bar s$ is system-specific and depends on the parameters of the quench. 

Even though the emergence of universality is apparently limited to $w \ll \bar{w}$, systems belonging to class B might display unexpected universal  
properties in the fully quantum regime $w>\bar{w}$.
In order to demonstrate this, we focus again on the free bosonic theory in Eq.~\reff{ham1} and we show that for $m_0 \rightarrow 0$ the statistics of the work displays a behavior analogous to the Bose-Einstein condensation of the ideal Bose gas in the grand canonical ensemble. This implies a transition in the large deviation statistics for $w>\bar{w}$ from exponential to algebraic. 
In fact, we note that the excess free energy 
$f_{ex}(s)$ in Eq.~\reff{eq:fex-HO} has the same form as half the scaled cumulant generating function $\psi(s)$ of the fluctuations of the spatial density $\rho_V$ of ideal Bose particles (of mass $m_B$) within a large region of volume $V$. At equilibrium in an %the grand canonical 
ensemble with  
chemical potential $\mu\le 0$ (in units of temperature $\beta^{-1}$) one finds
$\psi(s) = \int \frac{\rmd^dk}{(2\pi)^d} \ln \left(\frac{1-\Lambda_k \rme^{-s}}{1-\Lambda_k}\right)$,  where  $\Lambda_k = \rme^{-\beta\varepsilon_k+\mu}$ with $\varepsilon_k = \hbar^2 k^2/(2m_B)$ and the integral is over ${\mathbb R}^d$.
Accordingly the plot of $\psi(s)$ has the form B in Fig.~\ref{fig:LF}(a), with $\bar s = -\mu$.
%
%
%%%%%%%%%%%%%%%%%%%%%%%%%%%%%%%%%%%%%%%%%%%%%
%% expl. of Bose-Einstein condensation for the Bose gas
For the ideal Bose gas, the condensation occurs as $\mu \to \mu_c = 0$: %correspondingly, 
the asymptotic slope $\bar s = -\mu$ of the rate function $I(\rho >\bar \rho)$ vanishes together with the function itself (see Fig.~\ref{fig:LF}(b)). The mean value $\bar \rho = \langle \rho\rangle = \psi'(0)$ above which this happens is the critical density for condensation $\rho_c = l^{-d}\zeta(d/2)$  \cite{H-87}, which is finite only for $d>d_c=2$, where $l \equiv (2\pi\beta\hbar^2/m_B)^{1/2}$ is the thermal wavelength. 
$I$ vanishes for $\rho>\rho_c$ because the probability $p(\rho)$ acquires an algebraic dependence on $\rho$ --- due to the contributions of fluctuations in single-particle states with small $k$ --- and indeed 
the momenta $\langle (\rho-\bar\rho)^n \rangle$ with $n\ge d/2$  diverge as $\mu\to\mu_c$;
e.g., $\langle (\Delta\rho)^2 \rangle \propto (-\mu)^{-\alpha}$ where $\alpha = 2-d/2$ for $d<4$.
% and $\alpha=0$ for $d>4$.  
%

For the statistics of the work, $m_0$ plays a role similar to $\mu$, although 
the occupation  
of the energy levels is determined by the non-thermal distribution generated by the quench and not by the Bose statistics. % 
In fact, both $m_0$ and $\mu$  determine the $k$-dependence of $\lambda_{k\simeq 0}^2$ and $\Lambda_{k\simeq 0}$, respectively, on which the onset of the condensation depends. 
%%%%%%%%%%%%%%%%%%%%%%%%%%%%%%%%%%%%%%%%%%%%%
%% present case
In the case of the intensive work, $m_0$ is the control parameter: for $m_0\to 0$, $\langle w \rangle$ is finite for $d>d'_c=1$ with a corresponding "critical value"  $w_c(m)$.
The emergence of $d'_c \neq d_c$ is due to the fact that the dependence of $\lambda_{k\simeq 0}^2$ on $k$ crosses over from quadratic for $m_0\neq 0$ to linear for $m_0=0$. Analogous crossover occurs in %when studying 
the condensation of an ideal Bose gas with relativistic dispersion~\cite{BKM-79}. %$\varepsilon_k = \omega_k(m_0)$ 
The rate function $I(w)$ vanishes identically for $w>w_c$ and $p(w)$ acquires an algebraic dependence on $w$ because of the slow asymptotic decay of the probability distribution of the work done on modes with small $k$, which are mildly confined in the initial state with $m_0\to 0$.  As a result, moments $\langle (w-\bar w)^n\rangle$ with $n \ge d$ diverge in this limit with, e.g., 
%
%$\langle (\Delta w)^2\rangle \propto m_0^{-\alpha'}$ with $\alpha' = 2 - d$
%
$\langle (\Delta w)^2\rangle \propto \ln (m/m_0)$ in $d=2$. 

\emph{Conclusions --}  We discussed the qualitative features of the large deviation statistics of the work done during a quantum quench, highlighting the emergence of universality and, for bosonic systems, of a non-thermal condensation transition. Even though large fluctuations are exponentially rare as the system size increases, the value of the rate function $I(0)$ can be reduced by a suitable choice of the quench parameters, making them observable by a post-selection of experimental data. 

\emph{Acknowledgements --} AG and AS are grateful to KITP for hospitality. This research was supported in part by the National Science Foundation under Grant No. NSF PHY11-25915. AG is supported by MIUR within "Incentivazione alla mobilit\`a di studiosi stranieri e italiani residenti all'estero."


\begin{thebibliography}{99}
\bibitem{footnote1}
As discussed further below, this is not actually the case when a condensation transition occurs.
\bibitem{footnote2}
Here we assume that $I(w)$ vanishes only at a single point $w=\bar w$, so that the most probable value coincides with the mean value for $N\to\infty$ \cite{footnote1}.
\bibitem{NotaBene}
Note that the edge singularities discussed in Ref.~\cite{GS-11} 
are scaled onto the threshold when studied in term of the intensive work $w_N$.
\end{thebibliography}
\end{document}